\documentclass[prd,twocolumn,nopacs,nofootinbib]{revtex4}

\usepackage{amsmath}
\usepackage{graphicx}
\usepackage{amsfonts}
\usepackage{latexsym}
\usepackage{bbold}
\usepackage{wasysym}
\usepackage{calligra}
\usepackage{float}
\usepackage{ulem}
\usepackage{inputenc}
\usepackage{xspace}
\usepackage{url}
\usepackage{epstopdf}
\usepackage{tikz}
\usepackage{amssymb}
\usepackage{enumitem}

\usepackage{enumitem}
\usepackage{appendix}

\usepackage{amsmath}
\usepackage{amssymb}
\usepackage{graphicx}
\usepackage{amsfonts}
\usepackage{latexsym}
\usepackage{bbold}
\usepackage{wasysym}
\usepackage{calligra}
\usepackage{float}
\usepackage{ulem}
\usepackage{inputenc}
\usepackage{xspace}
\usepackage{url}
\usepackage{epstopdf}
\usepackage{tikz}
\usepackage{amsthm}
\usepackage{soul}

\newcommand{\be}{\begin{equation}}
\newcommand{\ee}{\end{equation}}
\newcommand{\beq} {\begin{equation}}
\newcommand{\eeq} {\end{equation}}
\newcommand{\ba}{\begin{eqnarray}}
\newcommand{\ea}{\end{eqnarray}}

\begin{document}

	\title{	On the role of the Parity Violating Hojman--Holst term in Gravity Theories}
	\author{Damianos Iosifidis$^{1,2}$}
   \affiliation{$^{1}$Scuola Superiore Meridionale, Largo San Marcellino 10, 80138 Napoli, Italy}
    
\affiliation{$^{2}$INFN– Sezione di Napoli, Via Cintia, 80126 Napoli, Italy}

	\email{d.iosifidis@ssmeridionale.it}

	\date{\today}
	\begin{abstract}
We study Parity Violating Gravity Theories whose gravitational Lagrangian is a generic function of the scalar curvature and the parity odd curvature pseudoscalar,  commonly known as the Holst (or Hojmann) term. Generalizing some previous results in the literature, we explicitly show that if the Hessian of this function is non-degenerate, the initial non-Riemannian Theory is on-shell equivalent to a metric Scalar-Tensor Theory. The generic form of the kinetic coupling function and the scalar potential of the resulting Theory are explicitly found and reported.
		
	\end{abstract}
	
	\maketitle
	
	\allowdisplaybreaks
	
	

\section{Introduction}
In Riemannian geometries there is only one scalar  one can construct (from contractions) that is linear in the curvature tensor and this is the Ricci scalar. However, if one allows for torsion, there is also a pseudoscalar quantity that can be formed in 4-dimensions, the pseudoscalar curvature $\varepsilon^{\alpha\beta\mu\nu}R_{\alpha\beta\mu\nu}$, commonly referred to as the Holst term \cite{Holst:1995pc}. However, it is important to stress that its inclusion in gravity theories was first considered by Hojman, Mukku and Sayed \cite{Hojman:1980kv} and  around the same time also by Nelson \cite{Nelson:1980ph}, some fifteen years prior to Holst.   Ever-since there has been an ever increasing interest with  several studies  focusing on  the matter, see e.g. \cite{Hecht:1996np,Banerjee:2010yn,Shapiro:2014kma,Langvik:2020nrs,Bombacigno:2018tih,BeltranJimenez:2019hrm,Shaposhnikov:2020gts,Obukhov:2020zal,Bombacigno:2021bpk,Pradisi:2022nmh,Salvio:2022suk,Gialamas:2024iyu,Racioppi:2024pno,DiMarco:2023ncs,Gialamas:2024cit,Gialamas:2024uar,Katsoulas:2025mcu} and in particular its possible importance for Inflation.

Ultimately the question to be answered is how many new degrees of freedom does the presence of this term bring into the game. 
There have been several studies discussing the role of this parity-violating term. 
 In particular, in \cite{BeltranJimenez:2019hrm} it was shown that in the quadratic Poincare gauge theory, the inclusion of the pseudoscalar term gives rise to an additional pseudoscalar degree of freedom.
 A generalization was given in \cite{Pradisi:2022nmh} where general functions $f(R,\mathcal{H})$ where considered but the dynamical equivalence to Scalar-Tensor was established only for special cases. It is our purpose here to generalize this result and establish the complete Scalar-Tensor equivalence by giving the exact forms of the kinetic function coupling and the potential of the resulting Theory. By using Inflation data one can then reconstruct the form of the function $f(R,\mathcal{H})$ that gives the desired results.

The paper is organized as follows. We firstly communicate the basic ingredients needed for this work. Then we study thoroughly generic $f(R,\mathcal{H})$ Theories, after classifying them according to the properties of the function $f$. For the case where the latter has a non-vanishing Hessian we explicitly obtain equivalence with Scalar-Tensor Theories both at the level of the equations of motion as well as to that of the action. We then conclude our results and point to future directions.

\section{Conventions/Notation}
Let us briefly go over the basic definitions and conventions that we shall be using throughout. We start with a 4-dimensional Metric-Affine space \cite{schouten1954ricci,hehl1995metric} consisting of a metric $g$ and a linear connection $\nabla$ whose components in local coordinates read, $g_{\mu\nu}$ and $\Gamma^{\lambda}{}_{\mu\nu}$ respectively. From these we define the curvature, torsion and non-metricity tensors according to
\begin{subequations}
\label{eq: spin only}
\begin{align}
R^{\mu}_{\;\;\;\nu\alpha\beta}&:= 2\partial_{[\alpha}\Gamma^{\mu}_{\;\;\;|\nu|\beta]}+2\Gamma^{\mu}_{\;\;\;\rho[\alpha}\Gamma^{\rho}_{\;\;\;|\nu|\beta]} \;\;,\label{R} \\
S_{\mu\nu}^{\;\;\;\lambda}&:=\Gamma^{\lambda}_{\;\;\;[\mu\nu]}
\;\;, \;\;  \\
Q_{\alpha\mu\nu}&:=- \nabla_{\alpha}g_{\mu\nu}
\end{align}
\end{subequations}
respectively. From the curvature we define the Ricci tensor as usual $R_{\mu\nu}:=R^{\lambda}{}_{\mu\lambda\nu}$, which now it is not symmetric in general,  and from it we derive the scalar curvature $R:=g^{\mu\nu}R_{\mu\nu}$ associated to the general connection. An important quantity that will concern us here, which is vanishing in Riemannian geometries but non-vanishing in the presence of torsion, is the \underline{pseudoscalar curvature}
\beq
\mathcal{H}:=\varepsilon^{\alpha\beta\mu\nu}R_{\alpha\beta\mu\nu}
\eeq
that frequently goes by the name 'Holst term' even though, as mentioned in the introduction, there were works even many years prior to Holst that considered this term \cite{Hojman:1980kv,Nelson:1980ph}. The latter along with the scalar curvature are the two scalar quantities that can be formed that are linear in the curvature.

The difference between the general connection and the Levi-Civita connection defines the distortion
\cite{schouten1954ricci,hehl1995metric}
\begin{gather}
N^{\lambda}_{\;\;\;\;\mu\nu}:=\Gamma^{\lambda}_{\;\;\;\mu\nu}-\widetilde{\Gamma}^{\lambda}_{\;\;\;\mu\nu}=\nonumber \\
\frac{1}{2}g^{\alpha\lambda}(Q_{\mu\nu\alpha}+Q_{\nu\alpha\mu}-Q_{\alpha\mu\nu}) -g^{\alpha\lambda}(S_{\alpha\mu\nu}+S_{\alpha\nu\mu}-S_{\mu\nu\alpha}) \label{N}
\end{gather}
where $\widetilde{\Gamma}^{\lambda}_{\;\;\;\mu\nu}$ is the usual Levi-Civita connection derived only from the metric and its first order derivatives. 

From torsion and non-metricity one constructs 3 vectors and 1 pseudovector according to 

\beq
S_{\mu}:=S_{\mu\lambda}^{\;\;\;\;\lambda} \;\;, \;\;\;
t_{\mu}:=\epsilon_{\mu\alpha\beta\gamma}S^{\alpha\beta\gamma}  
\eeq
\beq
Q_{\alpha}:=Q_{\alpha\mu\nu}g^{\mu\nu}\;,\;\; q_{\nu}=Q_{\alpha\mu\nu}g^{\alpha\mu}
\eeq
We may then decompose torsion and non-metricity as \cite{McCrea:1992wa,hehl1995metric,Obukhov:1997zd}
\begin{equation}
	S_{\mu\nu\lambda}=-\frac{2}{3} g_{\lambda[\mu}S_{\nu]}+\frac{1}{6} \epsilon_{\mu\nu\lambda\rho}\tilde{S}^{\rho}+Z_{\mu\nu\lambda}
	\end{equation}
 \beq
	Q_{\alpha\mu\nu}=\frac{\Big( 5 Q_{\alpha}-2 q_{\alpha} \Big)}{18}g_{\mu\nu}+\frac{\Big(  4 q_{(\mu}g_{\nu )\alpha}-Q_{(\mu}g_{\nu )\alpha}   \Big) }{9}+\Omega_{\alpha\mu\nu} \label{Qrepr}
	\eeq
	 respectively, where $Z_{\alpha\mu\nu}$  and $\Omega_{\alpha\mu\nu}$ are the traceless parts, satisfying 
   $ \epsilon^{\mu\nu\lambda\alpha}Z_{\mu\nu\lambda}=0 \;, $ $\;Z_{\mu\nu\lambda}g^{\nu\lambda}=0$ and
    $\Omega_{\alpha\mu\nu}g^{\mu\nu}=0 \;,\;$  $\Omega_{\alpha\mu\nu}g^{\alpha\mu}=0$.

Finally, for latter use we define the so-called Palatini tensor as\footnote{This tensor appears when varying the scalar curvature with respect to the general connection.}
\begin{gather}
	P_{\lambda}^{\;\;\;\mu\nu}=-\frac{\nabla_{\lambda}(\sqrt{-g}g^{\mu\nu})}{\sqrt{-g}}+\frac{\nabla_{\sigma}(\sqrt{-g}g^{\mu\sigma})\delta^{\nu}_{\lambda}}{\sqrt{-g}} 
	\nonumber \\+2(S_{\lambda}g^{\mu\nu}-S^{\mu}\delta_{\lambda}^{\nu}+g^{\mu\sigma}S_{\sigma\lambda}^{\;\;\;\;\nu}) = \nonumber \\
=\delta^{\nu}_{\lambda}\left( q^{\mu}-\frac{1}{2}Q^{\mu}-2 S^{\mu} \right) + g^{\mu\nu}\left( \frac{1}{2}Q_{\lambda}+2 S_{\lambda} \right)\nonumber \\-( Q_{\lambda}^{\;\;\;\mu\nu}+2 S_{\lambda}^{\;\;\;\;\mu\nu})
	\end{gather}
with traces $P^{\lambda}{}_{\lambda\nu}=0$, $P_{\mu}:=P^{\lambda}{}_{\mu\lambda}$ and $\tilde{P}^{\lambda}:=P^{\lambda}{}_{\mu\nu}g^{\mu\nu}$.  The fact that the first contraction of this tensor vanishes identically has to do with the projective invariance\footnote{Projective transformations are defined through the equivalence relation $\Gamma^{\lambda}{}_{\mu\nu}\rightarrow\Gamma^{\lambda}{}_{\mu\nu}+\delta^{\lambda}_{\mu}\xi_{\nu} $, with $\xi_{\nu}$ being an arbitrary 1-form. In general if a theory is projective invariant the $\Gamma$-variation of the gravitational action always produces a tensor with a vanishing first contraction \cite{Iosifidis:2019fsh}. Furthermore the field equations of such theories are also projective invariant and one can set the gauge $\xi_{\nu}$ at their will (for more details see \cite{Iosifidis:2019fsh}  and also \cite{Iosifidis:2018zwo}). In our case both $R$ and $\mathcal{H}$ have this symmetry and therefore we may gauge fix conveniently. } of the scalar curvature $R$. We now have all the necessary ingredients needed to start  our investigation.

\section{The role of pseudoscalar curvature in gravity}
As discussed in the introduction, our aim is to generalize some known results in the literature about the role of the peudoscalar curvature term in gravity. To this end, we consider the generic action
\beq
S=\frac{1}{2\kappa}\int d^{4}x \sqrt{-g}f(R,\mathcal{H})\label{Actionf}
\eeq
where $R=g^{\mu\nu}R_{\mu\nu}$ is the Ricci scalar and $\mathcal{H}=\varepsilon^{\alpha\beta\gamma\delta}R_{\alpha\beta\gamma\delta}$ pseudoscalar curvature term.  We want to investigate how many new degrees of freedom does this Theory have compared to General Relativity. Let us derive the associated field equations. Variation with respect to the metric yields
\beq
f_{R}R_{(\mu\nu)}-\frac{1}{2}g_{\mu\nu}(f-\mathcal{H}f_{\mathcal{H}})-f_{\mathcal{H}}\varepsilon_{(\mu}{}^{\alpha\beta\gamma}R_{\nu)\alpha\beta\gamma}=0\label{metricfe}
\eeq
while, varying with respect to the connection we get
\begin{gather}
f_{R}P_{\lambda}{}^{\mu\nu}+\delta_{\lambda}^{\nu}\partial^{\mu}f_{R}-g^{\mu\nu}\partial_{\lambda}f_{R}-2 \varepsilon_{\lambda}{}^{\mu\alpha\beta}S_{\alpha\beta}{}{}^{\nu}f_{\mathcal{H}}\nonumber  \\+   \frac{2}{\sqrt{-g}}(2 S_{\alpha}-\nabla_{\alpha})(\sqrt{-g}f_{\mathcal{H}}\varepsilon_{\lambda}{}^{\mu\alpha\nu})=0 \label{conf}
\end{gather}
or equivalently,
\begin{gather}
f_{R}P_{\lambda}{}^{\mu\nu}+\delta_{\lambda}^{\nu}\partial^{\mu}f_{R}-g^{\mu\nu}\partial_{\lambda}f_{R}+2 f_{\mathcal{H}}(S_{\alpha\beta\lambda}+Q_{\alpha\beta\lambda})\varepsilon^{\alpha\beta\mu\nu}\nonumber \\-2 f_{\mathcal{H}}S_{\alpha\beta}{}{}^{\mu}\varepsilon_{\lambda}{}^{\alpha\beta\nu}-2 \varepsilon_{\lambda}{}^{\mu\alpha\nu}\partial_{\alpha}f_{\mathcal{H}}=0 \label{confe}
\end{gather}
where we have used the definition of non-metricity to arrive at (\ref{confe}).
Raising one index and then contracting the latter with $\varepsilon_{\gamma\lambda\mu\nu}$ after some algebra it follows that
\beq
f_{R}t_{\mu}+(8 S_{\mu}+2 Q_{\mu}-2 q_{\mu})f_{\mathcal{H}}-6 \partial_{\mu}f_{\mathcal{H}}=0 \label{econtracted}
\eeq
Furthermore, taking the trace of (\ref{confe}) one time in $\lambda=\nu$ and another with $g_{\mu\nu}$, after renaming indices we get
\beq
f_{R}P_{\mu}+3 \partial_{\mu}f_{R}+2 f_{\mathcal{H}}t_{\mu}=0\label{fp}
\eeq 
and
\beq
f_{R}\tilde{P}_{\mu}-3 \partial_{\mu}f_{R}-2 f_{\mathcal{H}}t_{\mu}=0
\eeq
respectively. Adding them up it follows that
\beq
P_{\mu}+\tilde{P}_{\mu}=n q_{\mu}-Q_{\mu}=0
\eeq
Now recall the projective freedom. We may choose the gauge such that $q_{\mu}=0$. Then from the latter equation we have that $Q_{\mu}=0$ as well. Furthermore, for this gauge choice, $P_{\mu}=-\tilde{P}_{\mu}=-4 S_{\mu}$ and (\ref{fp}) simplifies to
\beq
-4 f_{R}S_{\mu}+3 \partial_{\mu}f_{R}+2 f_{\mathcal{H}}t_{\mu}=0
\eeq
Using these results we may then combine the latter with (\ref{econtracted}) in order to eliminate $t_{\mu}$ and we find
\beq
S_{\mu}=\frac{3}{4}\left( \frac{f_{R}\partial_{\mu}f_{R}+4 f_{\mathcal{H}}\partial_{\mu}f_{\mathcal{H}}}{f_{R}^{2}+4 f_{\mathcal{H}}^{2}}\right)
\eeq
Quite remarkably, this means that the torsion vector is exact since, obviously, the latter may also be recast as
\beq
S_{\mu}=\frac{3}{8}\partial_{\mu}\Big[\ln{(f_{R}^{2}+4 f_{\mathcal{H}}^{2})}\Big] \label{Sform}
\eeq
Note that for the special case $f_{\mathcal{H}}=\alpha=const.$ this is in perfect agreement with the result in \cite{Iosifidis:2020dck} (see eq. ($4.24$) there). Substituting this back to either (\ref{econtracted}) or (\ref{fp}) yields the torsion pseudo-vector\footnote{The last equality follows from direct differentiation of the identity $\arctan{x}+\arctan{\Big(\frac{1}{x}\Big)}=\pm \frac{\pi}{2}$ with the plus sign holding for $x>0$ and the minus for $x<0$.}
\beq
t_{\mu}=3 \partial_{\mu}\left[ \arctan{\left( \frac{2 f_{\mathcal{H}}}{f_{R}} \right)} \right]=-3 \partial_{\mu}\left[ \arctan{\left( \frac{f_{R}}{2 f_{\mathcal{H}}} \right)} \right] \label{tform}
\eeq
which is also exact and  again in agreement with \cite{Iosifidis:2020dck} for the special case of constant $f_{\mathcal{H}}$. Now using the torsion and nonmetricity decompositions along with the fact that both nonmetricity vectors are vanishing, the Palatini tensor takes the form
\beq
P_{\lambda\mu\nu} =\frac{4}{3}S_{\lambda}g_{\mu\nu}-\frac{4}{3}S_{\mu}g_{\nu\lambda}-\frac{1}{3}\varepsilon_{\lambda\mu\nu\alpha}t^{\alpha}-(\Omega_{\lambda\mu\nu}+2 Z_{\lambda\mu\nu})
\eeq
Substituting this expression into (\ref{confe}) and also using the relations (\ref{econtracted}) and (\ref{fp}) we find the constraint for the rest of the modes
\begin{gather}
-f_{R}(\Omega_{\lambda\mu\nu}+2 Z_{\lambda\mu\nu})+2 f_{\mathcal{H}}(\Omega_{\alpha\beta\lambda}+Z_{\alpha\beta\lambda})\varepsilon^{\alpha\beta}{}{}_{\mu\nu}\nonumber \\-2 f_{\mathcal{H}}Z_{\alpha\beta\mu}\varepsilon^{\alpha\beta}{}{}_{\lambda\nu}=0 \label{tensormodes}
\end{gather}
As we  explicitly show in the appendix, for real valued functions $f(R,\mathcal{H})$, taking contractions with the Levi-Civita tensor, the above equation demands that
\beq
Z_{\mu\nu\lambda}=0 \;\;, \;\; \Omega_{\mu\nu\lambda}=0
\eeq
namely, the tensor modes of both torsion and nonmetricity vanish. Given the fact that the nonmetricity vectors are also equal to zero, it follows from (\ref{Qrepr}) that the full nonmetricity tensor vanishes,
\beq
Q_{\alpha\mu\nu}=0
\eeq
We conclude therefore that in $f(R,\mathcal{H})$ Theories in vacuum, the nonmetricity tensor can be set to zero with an appropriate gauge choice. It carries no dynamics. As for torsion, only the vector and pseudo-vector pieces survive and the full torsion tensor is given by
\beq
S_{\mu\nu\lambda}=\frac{2}{3}S_{[\mu}g_{\nu]\lambda}+\frac{1}{6}\varepsilon_{\mu\nu\lambda\alpha}t^{\alpha}
\eeq
Consequently, the distortion tensor takes the form 
\beq
N_{\lambda\mu\nu}=\frac{4}{3}S_{[\mu}g_{\lambda]\nu}+\frac{1}{6}\varepsilon_{\mu\nu\lambda\alpha}t^{\alpha}
\eeq
With these at hand, we may then perform a post-Riemannian expansion on the scalar curvature and the parity violating pseudoscalar curvature term, to obtain, respectively, 
\beq
R=\tilde{R}-4\tilde{\nabla}_{\mu}S^{\mu}-\frac{8}{3}S_{\mu}S^{\mu}+\frac{1}{6}t_{\mu}t^{\mu} \label{scalpost}
\eeq
and
\beq
\mathcal{H}=2 \tilde{\nabla}_{\mu}t^{\mu}+\frac{4}{3}S_{\mu}t^{\mu} \label{Hpost}
\eeq
 Let us now turn our attention to the metric field equations. Taking the trace of the metric field equations (\ref{metricfe}), it follows that
\beq
R f_{R}+\mathcal{H}f_{\mathcal{H}}-2f=0 \label{treq}
\eeq
There are 3 possibilities for the latter equation:
\begin{enumerate}
    \item It has $k$ real  solutions and one can express $\mathcal{H}_{i}=F_{i}(R)$, with $i=1,2,...k$.
    \item Eq. (\ref{treq}) is identically satisfied and no relation between $\mathcal{H}$ and $R$ is obtained.
    \item It has no real solutions.
\end{enumerate}
Of course, the third possibility is of no physical interest, so from hereinafter we shall disregard it, assuming that only the first two possibilities are valid. The second possibility, namely the case where the trace equation (\ref{treq}) is identically satisfied, means that the form of $f(R,\mathcal{H})$ is such that the total action is conformally invariant. This occurs when the function $f$ is of the form
\beq
f=R^{2} G\Big( \frac{R}{\mathcal{H}}\Big)
\eeq
where $G\Big( \frac{R}{\mathcal{H}}\Big)$ is an arbitrary analytic function of the ratio $R/\mathcal{H}$. This basically means that $f$ is Euler-homogeneous of degree 2. A special case of the latter, is the quadratic Weyl gravity where $f=\alpha R^{2}+\beta \mathcal{H}^{2}+\gamma R \mathcal{H}$. This has been studied in detail in \cite{Karananas:2024xja,Karananas:2025xcv}\footnote{The situation is quite different in the metric case, see \cite{Hell:2023mph}.}. It is worth noting that in this case, the particle spectrum of the Theory contains also the massive graviton, even though the standard Einstein-Hilbert term $(\propto R)$ is absent from the action.

Let us now focus on the first possibility, namely functions for which $(\ref{treq})$ has real solutions. In order not to clutter the notation we shall assume that we work with one (of the possibly many) solution of the latter and write
\beq
\mathcal{H}=F(R) \label{HR}
\eeq
If we now set 
\beq
F_{1}(R)=\frac{f_{R}}{2f_{\mathcal{H}}}\;\;, \;\; F_{2}(R)=\frac{1}{2}\ln{(f_{R}^{2}+4 f_{\mathcal{H}}^{2})}
\eeq
\beq
t(R)=-\frac{3}{1+F_{1}^{2}}F_{1,R}\;\;,\;\; F_{3}(R)=2t_{,R}+t F_{2,R}
\eeq
and substitute expressions (\ref{Sform}) and (\ref{tform}) into (\ref{HR}), using also (\ref{Hpost}), after some trivial algebra we obtain
\beq
F_{3}(R)(\partial R)^{2}+2 t(R)\Box_{g}R=F(R)
\eeq
We clearly see that the latter describes the evolution of a new scalar mode $\phi=R$. Therefore, the Theory propagates an additional scalar degree of freedom, alongside the graviton. For specific forms of the function $f$ this fact was already quite well-known in the literature (see for instance \cite{Pradisi:2022nmh}). Here we proved this for generic forms of functions $f(R,\mathcal{H})$, which are not conformally invariant. Let us now proceed by establishing the formal equivalence with Scalar-Tensor Theories, by means of an on-shell equivalent action to (\ref{Actionf}).

\section{Establishing the equivalence to Scalar-Tensor}
For completeness, let us also establish the equivalence at the level of the  action.
We  consider two auxiliary fields $\chi$ and $\psi$ and write down the action
\beq
S=\frac{1}{2\kappa}\int d^{4}x \sqrt{-g}\Big[ f(\chi,\psi)+f_{\chi}(R-\chi)+f_{\psi}(\mathcal{H}-\psi) \Big] \label{equivact}
\eeq
Varying with respect to $\chi$ and $\psi$ we obtain a system of two equations whose solution gives
\beq
\chi=R \;\;, \;\; \psi=\mathcal{H}
\eeq
provided that
\beq
f_{\chi \chi}f_{\psi \psi}-f_{\chi \psi}^{2}\neq 0 \label{condition}
\eeq
Given that the last condition is satisfied, we can then plug the above expressions for the auxiliary fields back to (\ref{equivact}) and we obtain on-shell equivalence with (\ref{Actionf}). Now set 
\beq
f_{\chi}=\Phi \;\;, \;\; f_{\psi}=\Omega \label{fchi}
\eeq
and notice that on-shell we have that $f_{\chi}=f_{R}$ and $f_{\psi}=f_{\mathcal{H}}$ and therefore using the above, the trace equation (\ref{fchi}) implies that $\Omega=\Omega(\Phi)$. Therefore, we have only one extra field and this is $\Phi$. Then, inverting (\ref{fchi}) it follows that $\chi=\chi(\Phi,\Omega(\Phi))=\chi(\Phi)$ and $\psi=\psi(\Phi,\Omega(\Phi))=\psi(\Phi)$. With these we may equivalently express (\ref{equivact}) as
\beq
S=\frac{1}{2\kappa}\int d^{4}x \sqrt{-g}\Big[ \Phi R+\Omega \mathcal{H}-V(\Phi) \Big]\label{Action1}
\eeq
where we have defined the potential
\beq
-V(\Phi):=f(\chi(\Phi),\psi(\Phi))-\Phi\chi(\Phi)-\Omega(\Phi)\psi(\Phi) \label{pot}
\eeq
To proceed further, we substitute expressions (\ref{scalpost}) and (\ref{Hpost}) in (\ref{Action1}). After some trivial partial integrations the final form of (\ref{Action1}) reads
\beq
S=\frac{1}{2\kappa}\int d^{4}x \sqrt{-g}\Big[ \Phi \tilde{R}-\mathcal{K}(\Phi)(\partial \Phi)^{2}-V(\Phi) \Big]\label{ActionScalTen}
\eeq
with the kinetic term coupling given by
\begin{gather}
   - \mathcal{K}(\Phi)=3\frac{\Phi +8 \Omega \Omega_{\Phi}-4 \Phi\Omega_{\Phi}^{2}}{\Phi^{2}+4 \Omega^{2}}\nonumber \\+\frac{1}{(\Phi^{2}+4 \Omega^{2})^{2}}\Big[ -\frac{3}{2}\Phi(\Phi+4\Omega \Omega_{\Phi})^{2}\nonumber \\+6\Phi (\Phi \Omega_{\Phi}-\Omega)^{2}+6 \Omega (\Phi+4 \Omega \Omega_{\Phi})(\Phi \Omega_{\Phi}-\Omega)
   \Big] \label{kin}
\end{gather}
We see therefore that given the form of $f$ and the trace eq. (\ref{treq}, when invertible in $\chi=\chi(\Phi)$, we can find the explicit forms of the potential through (\ref{pot}) and kinetic coupling function through (\ref{kin}). Inversely, using inflationary scenarios as guides for the forms of the potential and kinetic function one can reconstruct the functional form of $f(R,\mathcal{H})$ that best fits observations.

\subsection{Example: Quadratic Theory}

Let us consider a concrete example. The most studied, straightforward and motivated choice is to go up to quadratic order in $R$ and $H$, namely take the function \cite{Gialamas:2022xtt,Karananas:2024xja} $f(R,\mathcal{H})=a_{1}R+a_{2}R^{2}+b_{1}\mathcal{H}+b_{2}\mathcal{H}^{2}+c_{1}R\mathcal{H}$, with the corresponding action
\beq
S=\frac{1}{2}\int d^{4}x \sqrt{-g}\Big[ a_{1}R+a_{2}R^{2}+b_{1}\mathcal{H}+b_{2}\mathcal{H}^{2}+\gamma R\mathcal{H}\Big] 
\eeq
In order to have the on-shell equivalence with (\ref{equivact}) the condition (\ref{condition}) must be satisfied, which requires $4 a_{2}b_{2}-\gamma^{2}\neq 0$. In this case, the trace equation (\ref{treq}) fixes\footnote{It is important to note that here $a_{1}$ and $b_{1}$  are non-vanishing. When both of them are equal to zero, the Theory is scale invariant and the trace equation (\ref{treq}) trivializes. In this case one can use this invariance to set one of the auxiliary fields to some constant value. For more details on the matter see e.g. \cite{Karananas:2024xja}. }
\beq
\mathcal{H}=-\frac{a_{1}}{b_{1}}R
\eeq
and consequently we find the on-shell relations
\beq
f=\left( a_{2}+b_{2}\frac{a_{1}^{2}}{b_{1}^{2}}-\gamma \frac{a_{1}}{b_{1}}\right) R^{2}
\eeq
\beq
\Phi=f_{R}=a_{1}+\Big( 2 a_{2}-\gamma \frac{a_{1}}{b_{1}}\Big) R
\eeq
\beq
\Omega=f_{\mathcal{H}}=b_{1}+\Big( \gamma- 2 b_{2} \frac{a_{1}}{b_{1}}\Big) R
\eeq
Now, if the condition
\beq
b_{1}\Big( 2 a_{2}-\gamma\frac{a_{1}}{b_{1}}\Big)=a_{1}\Big( \gamma-2 b_{2}\frac{a_{1}}{b_{1}}\Big) \label{constr}
\eeq
among the parameters is satisfied, then the function $\Omega(\Phi)$ reads $\Omega(\Phi)=\lambda \Phi$, with $\lambda=b_{1}/a_{1}$. It is then quite remarkable that the kinetic function (\ref{kin}) turns out to be independent of $\lambda$ and acquires the value
\beq
\mathcal{K}(\Phi)=-\frac{3}{2\Phi} \label{32}
\eeq
meaning that the Theory is actually a Brans-Dicke Theory with BD parameter $\omega_{0}=-3/2$. It is also very interesting that for this parameter space the potential vanishes identically, i.e. $V(\Phi)=0$ and the Theory is actually a prototype Brans-Dicke Theory \cite{Brans:1961sx} with the aforementioned BD parameter, i.e.
\beq
S=\frac{1}{2\kappa}\int d^{4}x \sqrt{-g}\Big[ \Phi \tilde{R}+\frac{3}{2 \Phi}(\partial\Phi)^{2}\Big] \label{BD}
\eeq
Quite remarkably,  without any further constraint, the parameters of the initial Theory disappear from the above on-shell equivalent action, meaning that for the whole parametric family the resulting Theory is always (\ref{BD})!  Without the constraint (\ref{constr}) and for generic values of the parameters, the potential is easily computed to be of the quadratic type,
\beq
V(\Phi)=C_{0}(\Phi-a_{1})^{2}\;\;, \;\;\;C_{0}=\frac{a_{2}\lambda^{2}+b_{2}-\lambda \gamma}{(2 a_{2}\lambda-\gamma)^{2}}
\eeq
and the kinetic function is given by
\begin{gather}
\mathcal{K}(\Phi)=-\frac{3}{2}\frac{(1+4 \nu^{2})\Phi+12\mu\nu}{\Big[ (1+4\nu^{2})\Phi^{2}+8 \mu \nu \Phi +4 \mu^{2}\Big]}\nonumber \\-\frac{6 \mu^{2}\Phi}{\Big[ (1+4 \nu^{2})\Phi+8 \mu \nu \Phi +4 \mu^{2}\Big]^{2}}
\end{gather}
where we have abbreviated
\beq
\mu=b_{1}-a_{1}\nu
\;\;, \;\;  \nu=\frac{\lambda \gamma-2 b_{2}}{2 \lambda a_{2}-\gamma}
\eeq
Note that $\mu=0$ corresponds to the condition (\ref{condition}) which reduces the above to (\ref{32}) as expected. To conclude, without any constraint among the parameters, the Quadratic MAG Theory (\ref{Actionf}), is on-shell equivalent to the Scalar-Tensor Theory with potential and kinetic function given by (\ref{pot}) and (\ref{kin}) respectively, in agreement with well known results.

\subsection{Note on the parity even subsector}

It is an obvious fact that if $\mathcal{H}$ enters the action only in even powers, then the underlying Theory would be parity even. This happens for functions of the form $f(R,\mathcal{H}^{2})$. Then, using the identities for the Levi-Civita tensor, it is trivial to obtain
\begin{gather}
\mathcal{H}^{2}=\varepsilon^{\alpha\beta\gamma\delta}\varepsilon_{\mu\nu\kappa\lambda}R_{\alpha\beta\gamma\delta}R^{\mu\nu\kappa\lambda} = \nonumber \\
=4 \Big( R_{[\mu\nu]\alpha\beta}+R_{\alpha\beta\mu\nu}+R_{\mu\alpha\beta\nu}\nonumber \\+R_{\nu\alpha\mu\beta}+R_{\beta\nu\mu\alpha}+R_{\alpha\mu\nu\beta}\Big) R^{\mu\nu\alpha\beta} \label{HRR}
\end{gather}
It is then a rather remarkable fact that for the action consisting of an arbitrary function of the scalar curvature and the above quadratic combinations of the curvature tensor, the additional degree of freedom is just a scalar mode. Therefore, the Theory
\beq
S=\int d^{4}x \sqrt{-g} f(R, \mathcal{H}^{2})
\eeq
propagates both a graviton and an additional scalar mode. Note that the combination of the curvature contractions in the last equality of (\ref{HRR}) can be formed for arbitrary dimension. It would then be quite interesting to see if the above result holds true for higher dimensions as well, namely that this combination still propagates only one additional scalar degree of freedom. Finally, it would also be worthwhile to see how the results of this study change when one adds quadratic torsion and non-metricity invariants to the action like the ones considered in \cite{Iosifidis:2024ndl} for the case of $f(R)$ Theories. There it was shown that an additional scalar degree of freedom propagates.  We expect that in the case of extended $f(R,\mathcal{H})$ Theories, the inclusion of these quadratic terms would introduce further degrees of freedom in addition to the one found here.

\section{Conclusions}

Generalizing existing results in the literature, we have considered the extended class of Gravitational Theories whose Lagrangian is an arbitrary function of the generalized scalar curvature $R$ and the parity violating pseudoscalar curvature $\mathcal{H}$, the so-called Hojman or Holst term. By adopting a Metric-Affine formulation\footnote{Namely, treating the metric and the connection as independent fields} and  by solving the corresponding connection field equations we explicitly showed that the initial Theory is on-shell equivalent to a specific Scalar-Tensor Theory that is metric and torsionless. The general expressions for the potential and the kinetic function have been reported.

It was already known in the literature that the inclusion of the pseudoscalar invariant   generally introduces a new pseudoscalar mode but this was only known for specific forms of the Lagrangian. Here we proved that it holds true for generic $f(R,\mathcal{H})$ Theories. The torsion in the initial non-Riemannian Theory is then mapped to the scalar mode in the resulting Scalar-Tensor Theory. In order for this equivalence to hold it is necessary that the Hessian of this function $f(R,\mathcal{H})$ is non-zero and also that the trace equation (\ref{treq}) has at least one real solution. It is worth mentioning that even for this generalized case, the non-metricity continues to be a non-dynamical field   that can be set to zero by an appropriate gauge choice of the projective freedom. In contrast, torsion plays an essential dynamical role in this construction that cannot be removed. As already mentioned, it would be interesting to see how the results of the current study will change if one adds to the $f(R,\mathcal{H})$ Lagrangian the quadratic invariants of torsion and non-metricity as in \cite{Iosifidis:2024ndl}. Finally, let us mention that our results can be straightforwardly generalized to functions $f(R,\mathcal{H},\psi)$, where $\psi$ is an additional scalar field, whose kinetic term and potential will also supplement  the action.

\section{Acknowledgements}

I would like to thank Friedrich Hehl for useful discussions and Ioannis Gialamas for bringing to my attention some important related references.

\section{Appendix}
Let us show here that for real $f(R,\mathcal{H})$, eq. (\ref{tensormodes}), which we express also here for convenience
\begin{gather}
-f_{R}(\Omega_{\lambda\mu\nu}+2 Z_{\lambda\mu\nu})+2 f_{\mathcal{H}}(\Omega_{\alpha\beta\lambda}+Z_{\alpha\beta\lambda})\varepsilon^{\alpha\beta}{}{}_{\mu\nu}\nonumber \\-2 f_{\mathcal{H}}Z_{\alpha\beta\mu}\varepsilon^{\alpha\beta}{}{}_{\lambda\nu}=0 \label{basic}
\end{gather}
implies that 
\beq
Z_{\mu\nu\lambda}=0 \;\;, \;\; \Omega_{\mu\nu\lambda}=0
\eeq
\textit{Proof.} Firstly, we start by setting 
\beq
X_{\mu\nu\lambda}:=\Omega_{\alpha\beta\lambda}\varepsilon^{\alpha\beta}{}_{\mu\nu}\;\;, \;\; Y_{\mu\nu\lambda}:=Z_{\alpha\beta\lambda}\varepsilon^{\alpha\beta}{}_{\mu\nu}
\eeq
which are by definition antisymmetric in the first pair of   indices. Using these, we may then contract (\ref{basic}) three times independently, with $\varepsilon^{\mu\nu\alpha\beta}$, $\varepsilon^{\lambda\nu\alpha\beta}$ and $\varepsilon^{\mu\lambda\alpha\beta}$, respectively. Collecting the resulting equations (with indices relabeled), 
we end up with the system:
\beq
-f_{R}(\Omega_{\lambda\mu\nu}+2 Z_{\lambda\mu\nu})+2 f_{\mathcal{H}}(X_{\mu\nu\lambda}+Y_{\mu\nu\lambda})-2 f_{\mathcal{H}}Y_{\lambda\mu\nu}=0
\eeq
\beq
4 f_{\mathcal{H}}\Omega_{[\alpha\beta]\nu}-f_{R}(X_{\alpha\beta\nu}+2 Y_{\alpha\beta\nu})=0
\eeq
\beq
f_{R}(-X_{\alpha\beta\nu}+2 Y_{\alpha\beta\nu})+4 f_{\mathcal{H}}Z_{\alpha\beta\nu}-4 f_{\mathcal{H}}\Omega_{[\alpha\beta]\nu}=0
\eeq
\beq
4 f_{\mathcal{H}}\Omega_{[\alpha\beta]\nu}+2 f_{\mathcal{H}}Z_{\alpha\beta\nu}-\frac{1}{2}f_{R}Y_{\alpha\beta\nu}=0
\eeq
Adding up the first two above it follows that $X_{\mu\nu\lambda}=2 f_{\mathcal{H}}Z_{\mu\nu\lambda}$. Then using the latter result we  add the last two equations from the above system and get
\beq
Z_{\alpha\beta\nu}=-\frac{3 f_{R}}{8 f_{\mathcal{H}}}Z_{\gamma\delta\nu}\varepsilon^{\gamma\delta}{}_{\alpha\beta} \label{Zeq}
\eeq
The latter implies that either $Z_{\mu\nu\lambda}=0$ or that $Z$ is self-dual: $Z_{\alpha\beta\nu}=\pm \frac{i}{2}Z_{\gamma\delta\nu}\varepsilon^{\gamma\delta}{}_{\alpha\beta}$. Of course the latter possibility is excluded since not only does it constrain the possible forms of the function $f$, it also demands that $f$ is complex valued. Therefore, the solution to (\ref{Zeq}), for arbitrary function $f$, reads 
\beq
Z_{\alpha\beta\nu}=0 \label{Z0}
\eeq
With this, equation (\ref{basic}) takes the form
\beq
-f_{R}\Omega_{\lambda\mu\nu}+2 f_{\mathcal{H}}\Omega_{\alpha\beta\lambda}\varepsilon^{\alpha\beta}{}_{\mu\nu}=0
\eeq
Taking the symmetric part in $\mu\nu$ and recalling that $\Omega_{\lambda(\mu\nu)}=\Omega_{\lambda\mu\nu}$ the above yields
\beq
\Omega_{\lambda\mu\nu}=0
\eeq
The latter equation together with (\ref{Z0}), therefore prove our statement.

\bibliographystyle{unsrt}
	\bibliography{ref}

\end{document}